# Zak phase and band inversion in dimerized one-dimensional locally resonant metamaterials

Weiwei Zhu[1], Ya-qiong Ding[1,2], Jie Ren[1,3,4], Yong Sun[1,*], Yunhui Li[1], Haitao Jiang[1], and Hong Chen[1,†]

[1] *MOE Key Laboratory of Advanced Micro-Structured Materials, School of Physics Science and Engineering, Tongji University, Shanghai 200092, China*

[2] *Science College, University of Shanghai for Science and Technology, Shanghai 200093, China*

[3] *Center for Phononics and Thermal Energy Science, Tongji University, Shanghai 200092, China*

[4] *Shanghai Key Laboratory of Special Artificial Microstructure Materials and Technology, Tongji University, Shanghai, 200092, China*

**Abstract:** Zak phase, which refers to the Berry's phase picked up by a particle moving across the Brillouin zone, characterizes the topological properties of Bloch bands in one-dimensional periodic system. Here the Zak phase in dimerized one-dimensional locally resonant metamaterials is investigated. It is found that there are some singular points in the bulk band across which the Bloch states contribute π to the Zak phase, whereas while in the rest of the band the contribution is nearly zero. These singular points associated with zero reflection are caused by two different mechanisms: the dimerization-independent anti-resonating of each branch, and the dimerization-dependent destructive interference in multiple backscattering. The structure undergoes a topological transition point in the band structure where the band inverts and the Zak phase, which is determined by the numbers of singular points in the bulk band, changes following a shift in dimerization parameter. Finally, the interface state between two dimerized metamaterial structures with different topological property in the first band gap is demonstrated experimentally. The quasi-one-dimensional configuration of the system allows one to explore topology-inspired new methods and applications in the sub-wavelength scale.

---

* yongsun@tongji.edu.cn
† hongchen@tongji.edu.cn

# I. INTRODUCTION

Inspired by the discovery of topological insulators,[1,2] electronic systems with nontrivial topology has attracted a lot of interests. The concept of topology has also been extended to photonic systems to realize exotic phenomena.[3,4] The topological properties of the bulk band can be characterized by topological invariants, which is proportional to the Berry phase picked up by a particle moving across the first Brillouin zone. In two dimensional systems, the topological invariant is first Chern number,[5] which determines the number of edge states in the band gap through bulk-boundary correspondence. In one dimensional (1D) system, the topological invariant is the Zak phase,[6] which plays an important role in understanding the polarization of polyacetylene,[7] and the topological edge state in one dimensional systems.[8-13]

Zak phase was firstly introduced to characterize the topological band of electron motions in 1D periodic solids.[6,7] Recently, Zak phase has been intensively investigated in various artificial 1D crystals, including optical lattices,[14] photonic lattices,[15,16] acoustic crystals,[17] photonic crystals,[18-21] metamaterials,[22,23] and so on. The relationship between the Zak phase of the bulk bands and the surface impedance of 1D photonic crystals in the gap has been revealed.[18] This "surface bulk correspondence" in 1D system can be used to determine the Zak phase in experiments.[19,20] In addition, it is found that the Zak phase of one isolated band is related to the resonance of the equally spaced scatters in acoustic locally resonant metamaterials.[23] Controlling the topological property of the first band gap of locally resonant metamaterials in a convenient manner is extremely important in sub-wavelength-scale wave manipulation. However, the first band of metamaterials composed of simple cells always contains one and only one resonance at zero frequency, which limit the freedom in adjusting the topological property of the first band gap.

In this work, we investigate the Zak phase in dimerized 1D locally resonant metamaterials composed of a backbone waveguide and side-coupled branches playing the role of scatters.[24,25] We show that the Zak phase of the system is determined by some singular points in the bulk band across which the Bloch states contribute $\pi$ to the Zak phase. These singular points associated with zero reflection states can be distinguished into two types. The one corresponding to the anti-resonance of the branch is independent on the dimerization, and the other one coming from the destructive interference in multiple backscattering is dependent on the dimerization. As the dimerization parameter changes continuously, our structure undergoes a topological phase transition in the band structure where the band inverts and the Zak phase, which is determined by the numbers of singular points in the bulk band, changes. In particular, dimerization can help us to generate one or two singular points in the first bulk band. Finally, the interface state in the first band gap of two dimerized metamaterials with different topology in the first bulk band is demonstrated experimentally. Dimerization in 1D locally resonant metamaterials may pave a new way to manipulate wave in sub-wavelength scale by topological phase transition.

The paper is organized as follows. In Sec. II, we systematically study the Zak phase and band inversion in dimerized 1D locally resonant metamaterials in theory. In particular, we find the topological property of the first band can be changed by tuning the dimerization parameter. Then the band inversion and the topological interface state are experimentally demonstrated in Sec. III. Finally, we conclude in Sec. IV.

## II. ZAK PHASE AND BAND INVERSION

The scheme of the dimerized 1D locally resonant metamaterial is shown in Fig. 1. It is composed of a backbone waveguide along which side branches are grafted periodically. Here the side branches are made of same materials as the backbone, and play the role of the local scatters providing a phase/amplitude change with nearly zero optical thickness. The unit cell is composed of a backbone waveguide with the periodic length $\Lambda = 2L$, and two symmetrically arranged branches with the same length $l_0 = L/2$, as shown in Fig. 1(a). A dimerization parameter $\Delta = (l-L)/L$ is introduced, in which $l$ is the distance between two branches in the same unit cell. As a comparison, the structure without dimerization ($\Delta = 0$), as given in Fig. 1(b), is also investigated in the following.

It is worth noting that the dimerized 1D chains, *i.e.* Su-Schrieffer-Heeger (SSH) model, have been widely studied with tight binding approximation.[26,27] Different from SSH model in which the resonators are near-field coupled, the branches in our structure can only interact with others through wave scattering. Therefore, our theoretical investigation begins with transfer matrix method, which allows one to obtain the band structure and reflection coefficient, with the help of Bloch theory.[28-30]

The transfer matrix for the branch determined by its length $l_0$ and effective refractive index $n_e$ can be written as

$$T_c = \begin{pmatrix} 1+\dfrac{j\tan kl_0}{2} & \dfrac{j\tan kl_0}{2} \\ -\dfrac{j\tan kl_0}{2} & 1-\dfrac{j\tan kl_0}{2} \end{pmatrix}, \quad (1)$$

where $j$ is the imaginary unit, $k = \omega n_e/c$ is the wave vector in the branch at a given frequency $\omega$ ($c$ is the speed of light in vacuum). The transfer matrix for an entire unit cell in Fig. 1(a) can be determined as $T = T_{l_1} T_c T_{l_2} T_c T_{l_1}$, where

$$T_{l_i} = \begin{pmatrix} e^{jkl_i} & 0 \\ 0 & e^{-jkl_i} \end{pmatrix} \quad (2)$$

is the transfer matrix for the backbone waveguide with length $l_1 = L - l/2$, $l_2 = l$. For a finite structure with $N$ unit cells, the total transfer matrix can be obtained from the product of transfer matrix for single unit cell, $T^{(N)} = T^N$. Then transmission and reflection coefficients of the structure can be expressed in terms of the total transfer matrix elements as follows,

$$t = \frac{1}{T_{22}^{(N)}}; \quad r = -\frac{T_{21}^{(N)}}{T_{22}^{(N)}} \quad (3)$$

We first calculate the reflection coefficient for the semi-infinite system with the help of impedance boundary condition.[31] The reflectivity and the reflection phase as a function of $\Delta$ and $\omega$ are shown in Figs. 2(a) and 2(b). Four topological phase transition points are indicated by white circles in Fig. 2(a). By tuning the dimerization $\Delta$, the band gaps close and reopen at those points, and the signs of the reflection phase in the band gaps invert as shown in Fig. 2(b). It should be noted the frequencies with zero reflectivity form several curves passing through those transition points.

The band structure of dimerized metamaterials shown in Fig. 1(a) can be obtained from the transfer matrix of the single unite cell with $\cos q\Lambda = \frac{T_{11}+T_{22}}{2}$, namely

$$\cos q\Lambda = \cos k\Lambda - \frac{\tan^2 kl_0}{4}\cos k\Lambda - \tan kl_0 \sin k\Lambda + \frac{\tan^2 kl_0}{4}\cos 2k(L-l) \quad (4)$$

Where $q$ is the Bloch wave vector, and $\Lambda = 2L$ is the periodic length of the unit cell. In the following, five cases with parameters $\Delta = -0.8, -0.68, -0.5, 0, -0.5$ are studied in detail. Corresponding band structures are calculated, as shown in Fig. 3. In Figs. 3(a), 3(c) and 3(e), all bands are isolated. Here we mark the first three bands with B1, B2 and B3, the first three band gaps with G1, G2 and G3, respectively. When $\Delta$ is increasing from -0.8 to -0.5, B2 and B3 touch each other around $\Delta = -0.68$ and then re-separate. The symmetry of the band edge state is analyzed and presented in Fig. 3 with the red dots for even symmetry and pink stars for odd symmetry. The symmetry of two band edge states in B2 and B3 besides G2 is exchanged. This means a topological phase transition and the transition point is $\Delta = -0.68$ in Fig. 3(b).[18] Another topological phase transition point at $\Delta = 0$ in Fig. 3(d) is confirmed by the exchange of symmetry of the band edge states besides G1, as shown in Fig. 3(c) and 3(e).

Photonic band inversion would lead to photonic topological insulator. It has been argued that the photonic topological trivial and non-trivial insulators can be described by two kinds of single negative materials ($\varepsilon_{eff}\mu_{eff} < 0$, where $\varepsilon_{eff}$ and $\mu_{eff}$ are effective permittivity and permeability), namely $\varepsilon_{eff}$-single negative (ENG) materials and $\mu_{eff}$-single negative (MNG) materials.[21,22,32] For the dimerized metamaterials considered above, the effective parameters $\varepsilon_{eff}$ and $\mu_{eff}$ in the gap frequency range are retrieved by using the standard method,[32,33] as given in Fig. 3 with orange squares and purple triangles, respectively. One can see that for cases of $\Delta = -0.8$ and $\Delta = -0.5$, the metamaterials in G2 are MNG and ENG, respectively. So from $\Delta = -0.8$ to $\Delta = -0.5$, the topological properties of G2 changed, which is consistent with the result from the band inversion analysis. Similarly, for cases of $\Delta = -0.5$ and $\Delta = 0.5$, the metamaterials in G1 are ENG and MNG, respectively. For the wave incident from vacuum, the reflection phase is positive $(0 \sim \pi)$ for the ENG medium and negative $(-\pi \sim 0)$ for the MNG medium. The reflection phase calculated from the semi-infinite systems and the effective parameters retrieved from the transmission/reflection coefficients of single unit cell are consistent with each other.

The topological properties of the band structure of 1D system can be characterized by Zak phase, which is used to characterize the topological energy bands of electron motions in one-dimensional periodic solids,[6,7] and then is extended to characterize the topological properties of one dimensional photonic systems with mirror symmetry unit cells.[18] Basing on the original definition of the Zak phase, one can calculate it as follows. First, discretization of the first Brillouin zone is made with $\delta q = \pi/10$. Then the phase difference between states $u_{n,q}$ and $u_{n,q+\delta q}$ can be written as:

$$\delta \varphi = -\mathrm{Arg}\left[ \int_{-\Lambda/2}^{\Lambda/2} dx u_{n,q}^*(x) \partial_q u_{n,q}(x) \right], \quad (5)$$

whose sum over the entire Brillouin zone is the Zak phase with

$$\theta_n^{Zak} = \sum_{BZ} \delta \varphi_{n,q} \mod 2\pi. \quad (6)$$

The Zak phase of the isolated bulk bands for different $\Delta$ is given in Figs. 3(a), 3(c), and 3(e). According to the "surface bulk correspondence" in 1D system, the sign of the reflection phase in the band gap is related to the sum of the Zak phase of all bands below the band gap.[18] In Fig. 3, the reflection phases in G2 for cases of $\Delta = -0.8$ and $\Delta = -0.5$ have opposite signs. The reflection phase in G1 for $\Delta = -0.5$ and $\Delta = 0.5$ also have opposite signs. In addition to the direct calculation, the Zak phase of individual band can also be determined by the symmetry of its band edge states at Brillouin zone center and boundary, which is 0 when they have the same symmetry and $\pi$ if they have different symmetry. We can see that the Zak phase obtained from the symmetry of band edge state is consistent with the Zak phase calculated from the direct calculation.

The phase difference $\delta\varphi_{n,q}$ between states $u_{n,q}$ and $u_{n,q+\delta q}$ as a function of $q$ for $\Delta = -0.8, -0.5, 0.5$ is plotted in Figs. 4(a), 4(b), and 4(c), respectively. One can see clearly that $\delta\varphi_{n,q}$ contribute zero to the Zak phase in most area of the first Brillouin zone, except some singular points (SPs) at which $\delta\varphi_{n,q}$ contribute $\pi$ (or $-\pi$). So the Zak phase of the band is 0 ($\pi$) if it contains even (odd) number of such SPs. Actually the energy of those SPs correspond to the resonance frequencies of the single unit cell, at which there is zero reflection. Evolution of SPs with dimerization parameter $\Delta$ is visualized by the group of curves marked by black dash in Figs. 2(a) and 2(c), where the reflection of corresponding structure is zero. With the increase of $\Delta$, the frequency of SPs become lower and lower, which makes SPs enter the lower band after the topological phase transition point. In Figs. 2(b) and 2(d), one can further find there is always a $\pi$-phase jump in reflection across these curves. It means in addition to the transitions of the surface impedance in the gap, there are also similar transitions in the band, *namely* the impedance of the dimerized 1D metamaterials on two sides of those curves are distinguished, either greater or smaller than that of the free waveguide.

For our systems, there are two kinds of SPs. One is for the anti-resonance frequency of the side branch, which means no scattering wave from the each branch. These singular points are determined by $\tan kl_0 = 0$. We call these singular points as no-scattering-induced singular point (NS-SP). In Fig.4, we see the 1$^{st}$ NS-SP is in B1 at $q = 0$, whose frequency is 0. The other is due to the destructive multiple scattering, *namely* the perfect destructive interference happen between the back scattering waves from the first branch and the second branch in one unit cell. We call it multiple-scattering-induced singular point (MS-SP). Frequencies of MS-SPs are dependent on the dimerizaion parameter $\Delta$, and can be determined by $\tan kl_0 \tan kl_2 = 0$. With tuning the dimerization $\Delta$, we can adjust MS-SP into different bands (as shown in Fig. 4) to achieve different topological materials. Especially, the MS-SP can be introduced into the first band in our system, which makes the topological phase transition be possible for the first band. This point is very important for wave manipulation using topological excitations in sub-wavelength scale.

## III. EXPERIMENTALLY DEMONSTRATION OF BAND INVERSION AND TOPOLOGICAL INTERFACE STATE

Microwave experiments are conducted to demonstrate the previous theoretical results on Zak phase and band inversion in dimerized 1D locally resonant metamaterials. Using printed-circuit-board technology two samples are fabricated on a 1.6-mm thick FR4 substrate (relative permittivity $\varepsilon_r = 4.75$) with a periodic length of $\Lambda = 56\,\text{mm}$, as shown in Figs. 5(a), and 5(b). They are composed of five unit cells with dimerization parameters $\Delta = -0.5$ and $\Delta = 0.5$, respectively. Corresponding measurements (red line) of reflection amplitude $|r|$ and phase $\text{Arg}(r)$, as well as calculations from the transfer matrix (blue), are given in Figs. 5(b), 5(c), 5(e) and 5(f). One can find for both of them there are three high-reflection frequency bands below 3.5 GHz, which are consistent with the calculated gaps (gray area). The gap positions of two samples are same, but the reflection phase in the same gap may be different. In detail, measured reflection phase in G1 is positive for the first sample [Fig. 5(c)], but negative for the second [Fig. 5(f)]. Similar results exist in G3. These phase difference demonstrate the topological phase transitions with respect to the band structure.

Moreover, the reflection spectra provide direct evidence of the singular points in bulk bands. By comparing Fig. 5(b) and Fig. 5(e) carefully, we find the average reflection amplitudes of two samples are very different in bulk bands. In B1, refection of the second sample is much lower than that of the first sample. As we discussed in Fig. 4, B1 of the second sample contains an additional MS-SP, which is a zero reflection state. Therefore, the lower average reflectivity in an isolated bulk band means that there is an additional SP in the corresponding band. The difference between two samples in the measured average reflectivity in B2, B3, and B4 are also consistent with the theoretical calculated SPs in Fig.4.

Finally, as an example of the topological excitations in sub-wavelength scale, the interface state in a pair structure is investigated in the third sample. As shown in Fig. 6(a), it contains two parts: the left half is composed of two unit cells with $\Delta = 0.5$, and the right half with $\Delta' = -0.5$. As discussed above, the Zak phases of the first band of two metamaterials with dimerization parameters $\Delta = 0.5$ and $\Delta' = -0.5$ are different. According to "surface bulk correspondence" in 1D system, there should be an interface state in their pair structure in the first band gap, which is protected by the topological phase transition across the interface. This interface state is demonstrated with a transmission peak at 1.13 GHz which is located in the original G1 area, as shown in Fig. 6(b). Measured field distribution in the backbone waveguide at 1.13 GHz is given in Fig. 6(c) with red dots, together with calculations (blue line). It clearly shows that the field is enhanced near the interface (domain wall) due to the interface state. It should be noted that the whole length of the sample is 224 mm, which is less than the wavelength (~265 mm) in vacuum at 1.13 GHz. One can design more compact structure by increasing the length of the branch with constant periodic length.

## IV. CONCLUSION

In conclusion, we theoretically and experimentally investigate the Zak phase and band inversion in dimerized 1D locally resonant metamaterials. We demonstrate that the Zak phase of the isolated bands mainly comes from some SPs associated to zero reflection. Those SPs are classified into two types triggered by different mechansims: NS-SP comes from the dimerization-independent anti-resonance of each branch, and MS-SP comes from the dimerization-dependent destructive interference in multiple backscattering. Specially, the MS-SP can be tuned into the first band by changing the dimerization, which makes it possible to change the topological properties of the first band gap. Those results are confirmed by microwave experiments, and a subwavelength interface state is realized by connecting two dimerized structure with different topological properties in the first band gap. The results in our work can be extended into more complex 1D system like trimerization, as well as 1D systems in acoustics, phononics and plasmonics.

## ACKNOWLEDGMENTS

This research was supported by National Key Research Program of China (2016YFA0301101), National Natural Science Foundation of China (NSFC) (11504236, 11674247, 11775159, 11474220), Natural Science Foundation of Shanghai (No. 17ZR1443800), and the Fundamental Research Funds for the Central Universities.

**Figure Captions**

FIG. 1. The scheme of the 1D locally resonant metamaterials made of a backbone waveguide and branch resonators. (a) One unit cell with the dimerization of $\Delta = (l-L)/L$. Here the symmetrical unit cell has a length of $\Lambda = 2L$. The distance between two identical branches in one unit cell is $l = (1+\Delta)L$. $l_0 = L/2$ is the length of the branch. The backbone waveguide and branch are made of same material. (b) Structure without dimerization ($\Delta = 0$)

FIG. 2. Reflection coefficient for the semi-infinite systems (a-b) and single unit cell (c-d) as a function of the frequency and the dimerization. (a) and (c) are the amplitude of the reflection coefficient. (b) and (d) are the reflection phase. Four topological phase transition points are indicated by the white circles. The states with zero reflection are marked by a group of black dashed curves.

FIG. 3. Band structures for 1D metamaterials with dimerization $\Delta = -0.8, -0.68, -0.5, 0,$ and $0.5$. The frequency domain is divided into bulk bands, and band gaps (grey area). For the convenience in expression, the $n^{th}$ isolated pass bands (band gaps) are named with green letters Bn (brown letters Gn). The effective constitutive parameters $\epsilon_{eff}$ and $\mu_{eff}$ of metamaterials in the band gaps are drawn with orange squares and purple triangles, respectively. In (b) and (d), band crossing occurs, whereas in (a), (c), and (e), all of the bands considered are isolated. Symmetry of the band edge state is indicated with red dot for even symmetry and pink star for odd symmetry. The Zak phases (0 or $\pi$) of the isolated bands are also given.

FIG. 4. Phase difference $\delta\varphi_{n,q}$ between states $u_{n,q}$ and $u_{n,q+\delta q}$ with $\delta q = 2\pi/20$ as a function of $q$ in the first Brillouin zone under different dimerization parameters: (a) $\Delta = -0.8$, (b) $\Delta = -0.5$, (c) $\Delta = 0.5$. Two types of singular points are indicated by NS-SP and MS-SP, respectively. $1^{st}(2^{nd})$ means the principle value (the second branch value) of the multivalued equations.

FIG. 5. Experiements of the dimerized 1D locally resonant metamaterials. (a) Photography of a sample with $\Delta = -0.5$. (b-c) Amplitude and phase of reflection coefficient of the sample shown in (a). The red lines are measurements, and the blue lines are theoretical calculations from the transfer matrix. (d-e) the sample with $\Delta = 0.5$, and corresponding results. The samples are fabricated on FR4 double-sided copper-clad boards, with the geometric parameters given in the figure.

FIG. 6. Experiments of the interface state with the topological transition in the first band. (a) Photography of the sample composed by two topology-distinct 1D dimerized metamterials, *i.e.* structures with $\Delta = 0.5$ and $\Delta = -0.5$. The interface is denoted by the dash line. (c) Measured (red) and calculated (blue) transmission spectra. There is a transmission peak at 1.13 GHz emerging in the first band gap, which is represented by the grey area. (c) Electric field distribution of the topological interface state. The measurements (red dots) are consistent with theoretical calculations (blue lines) very well.

**Figures**

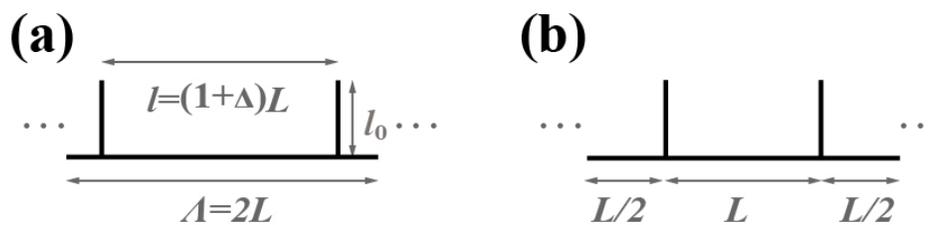

Fig. 1

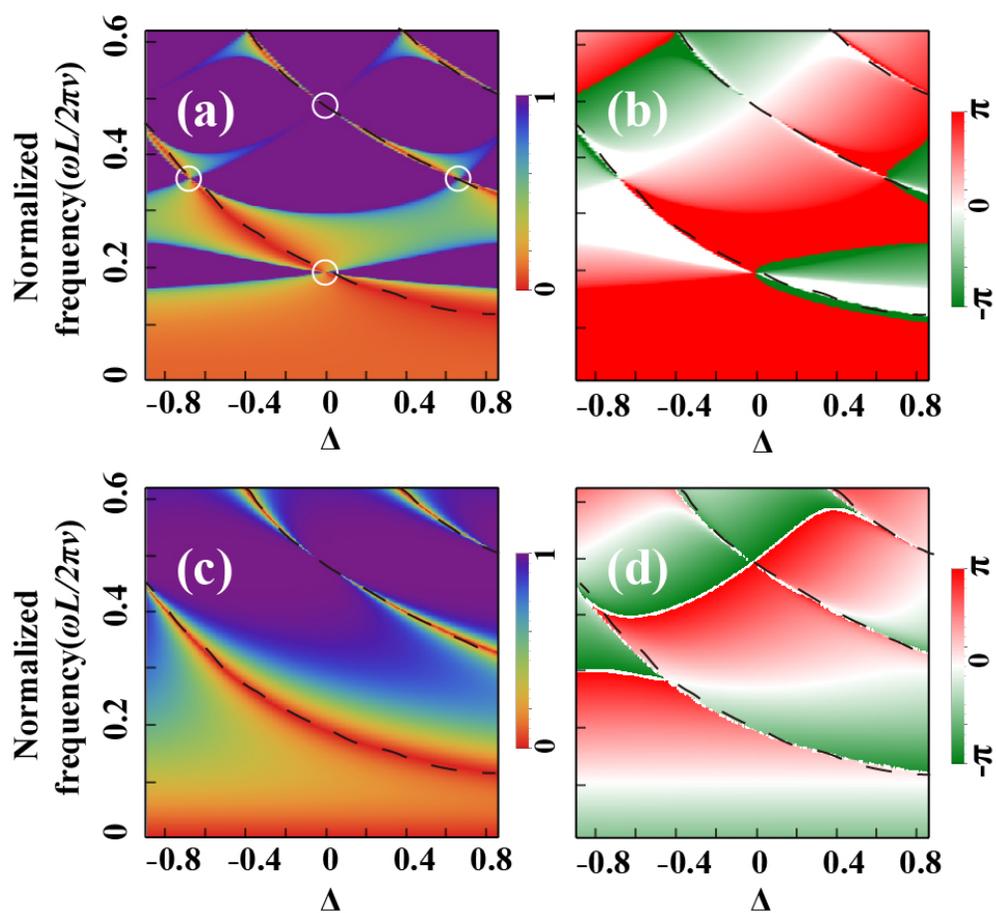

Fig. 2

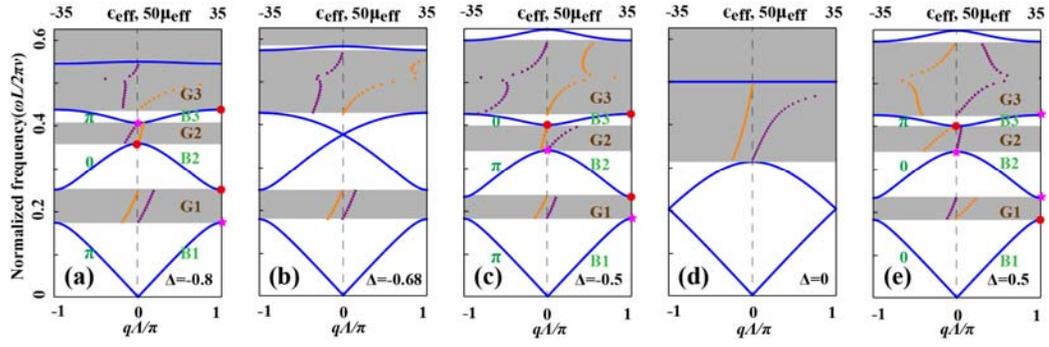

Fig. 3

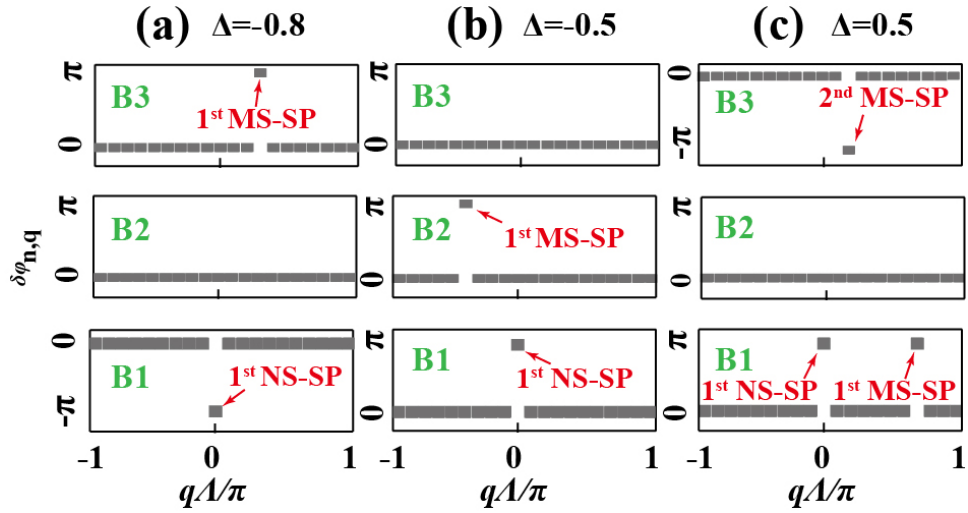

Fig. 4

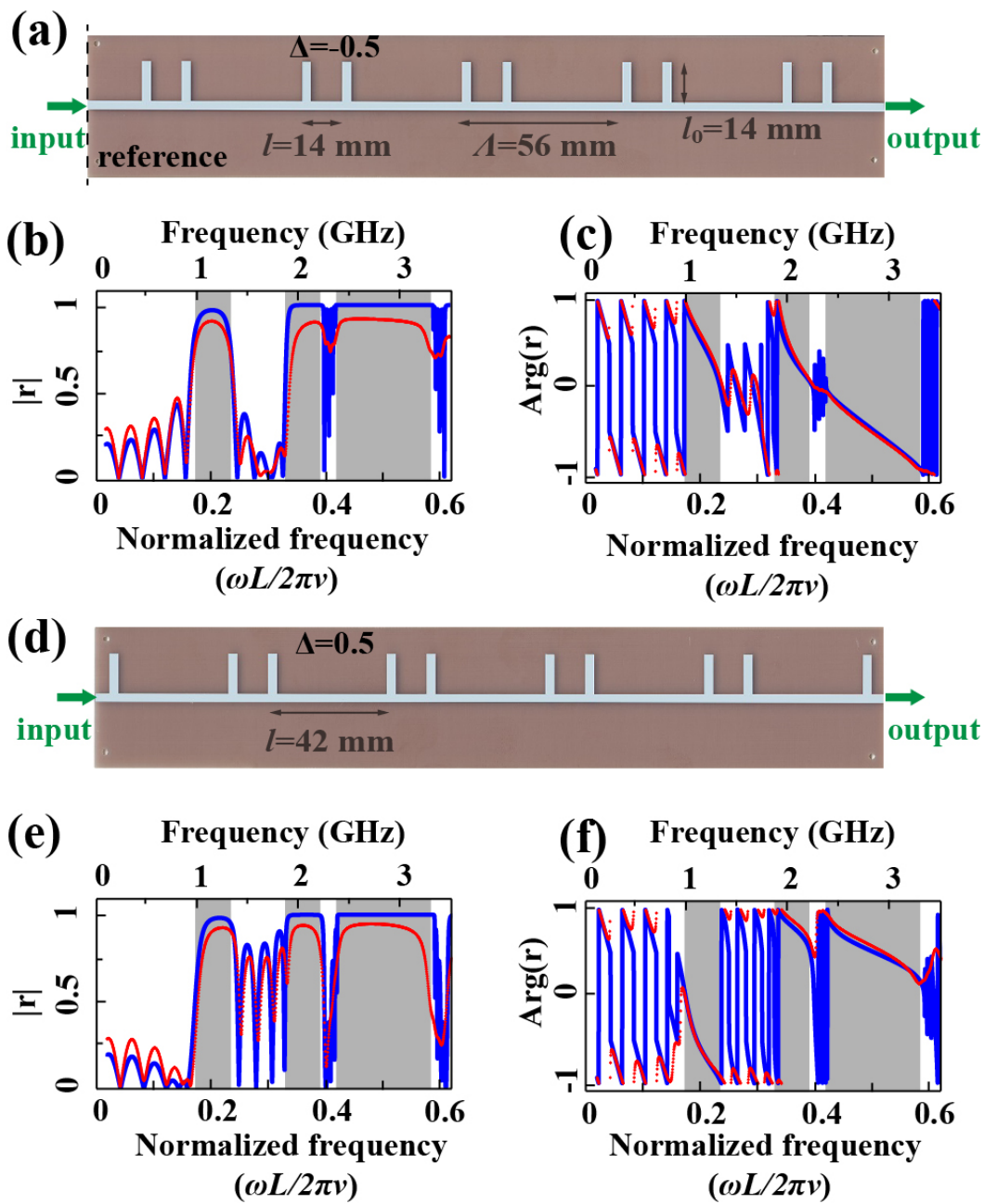

Fig. 5

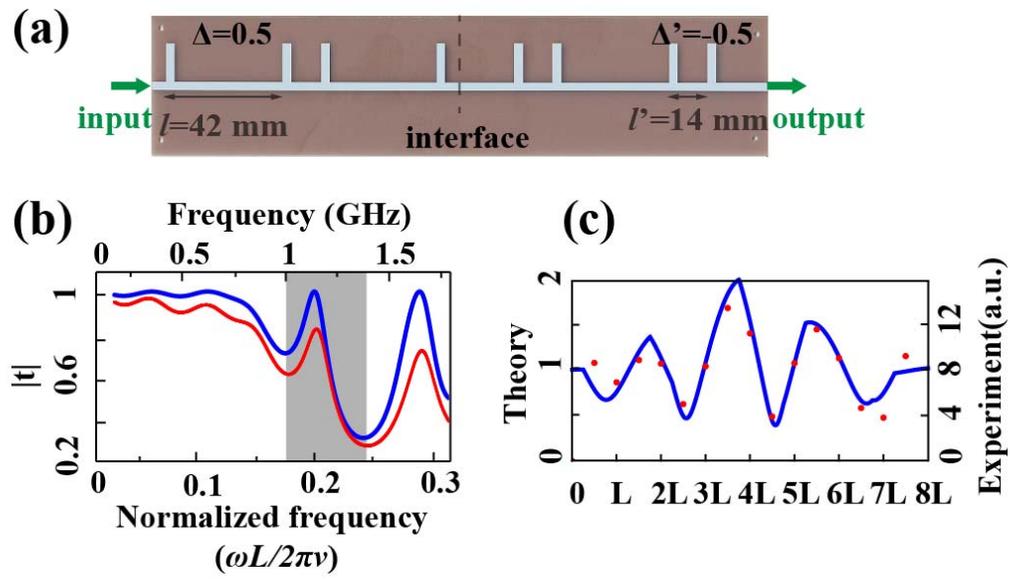

Fig. 6